# Density-functional theory of elastically deformed finite metallic system: work function and surface stress

V.V. Pogosov  and  V.P. Kurbatsky

[1)]Department of Microelectronics, Zaporozhye National Technical University, Zhukovski Str. 64, 69063  Zaporozhye, Ukraine

## Abstract

The effect of external strain on surface properties of simple metals is considered within the modified stabilized jellium model. The equations for the stabilization energy of the deformed Wigner-Seitz cells are derived as a function of the bulk electron density and the given deformation. The results for surface stress and work function of aluminium  calculated within the self-consistent Kohn-Sham method are also given. The problem of anisotropy of the work function of finite system is discussed. A clear explanation of independent experiments on stress-induced contact potential difference at metal surfaces is presented.

E-mail: vpogosov@zstu.zaporizhzhe.ua

## 1. INTRODUCTION

The early experimental investigations of the force acting on electron and positron inside a metallic tube in the earth's gravitational field [1,2] posed a question about the influence of metal deformation upon the electronic work function. The direct measurements, using the Kelvin method, showed a decreasing/increasing of the contact potential difference (CPD) of the tensed/compressed metal samples [3-5]. Similarly, the experiment with a high speed spinning metal rotor, which is nonuniformly deformed over the length, demonstrated that CPD changes between areas of the surface subjected to different deformation [6] (see also discussion of earlier experiments by Harrison [7]). The influence of deformation upon electronic emission from a thin metallic film has been also investigated [8]. Recently, a similar effect on CPD was observed at the surface of sample with a nonuniform distribution of residual mechanical stress [9]. These, at first sight, surprising results mean the respective increase/decrease of work function upon uniaxial tension/compression of metalic sample. Therefore, all these experiments stimulate two important questions which should be answered by the microscopic theory: (i)  Does the change of the CPD correspond to a change in the work function? (ii) Which sign will have the deformation gradients of surface energy and work function for a metal that is subject to the tension (or compression) along some direction?

The first question is connected with breaking of the local electroneutrality of metal and, as a result of this, with non-equipotentiality of its geometrical surface. The second question stems from the general statement of the theory of elasticity: the change in the total energy of a solid is proportional to the square of the relative deformation. Therefore, the energy must increase so for compression as for tension. On the other hand, experimentally it was found that in the range of elastic deformation, an uniaxial deformation of metallic sample leads to a linear change in the CPD [4,5]. This implies that classical theory of elasticity is not completely right in determination of elastic characteristics of surfaces. Besides, this question is of importance by determination of the surface tension or surface stress for macroscopic samples [10] and small metal particles [11].

The measurements of the derivative of surface tension of a solid with respect to the electrical variable (so-called an``estans'' [12]) indirectly show a little difference between the surface stress and surface energy . On the other hand, different calculations [13-15], including the ones based on the first principles [16], show the appreciable difference between these two quantities. A rough estimation of the difference between surface energy and surface stress can be also done using the cohesive energy and the vacancy formation energy. In the continuum approximation the cohesive energy (or atomic ``work function'') $\varepsilon_{coh}$ and the vacancy formation energy $\varepsilon_{vac}$ give respectively, the irreversible and reversible work required for creation of new spherical surface of Wigner-Seitz cell of radius $r_0$. Following [17]

$$\varepsilon_{coh} \approx 4\pi r_0^2 \gamma_0 \left(1 + \delta / r_0\right),$$

where $\gamma_0$ is surface energy per unit area of flat surface, $\delta / r_0$ the size correction for a surface of positive curvature. The reversible work for creation of vacancy (which can be defined as the work needed for blowing a small bubble) has the following form [18]

$$\varepsilon_{vac} \approx \int\limits_0^{r_0} dr \, 4\pi r^2 \left[2\tau_0 \left(1 - \delta / 2r\right)/ r\right] = 4\pi r_0^2 \tau_0 \left(1 - \delta / r_0\right).$$

Here we introduce a well defined physical quantity -- surface stress of flat surface, $\tau_0$ - to describe tensed curved surface [19,20]. Combining the expressions for $\varepsilon_{coh}$ and $\varepsilon_{vac}$, we obtain

$$\tau_0 \approx \gamma_0 \left(\frac{1 + \delta / r_0}{1 - \delta / r_0}\right) \frac{\varepsilon_{vac}}{\varepsilon_{coh}},$$

The Kohn-Sham calculations in Refs.[21,22] give $\delta/r_0 \approx 0.40$ and 0.52 for Na and Al, and the ratio of the experimental values $\varepsilon_{vac}/\varepsilon_{coh}$ is approximately equal or less then 1/3, respectively. These values agree very well with $\delta/r_0 \approx 1/2$ obtained in Ref.[18] which follows from Langmuir semi-empirical rule [23]. From this simple estimation follows that $\tau_0$ is approximately equal or less than $\gamma_0$.

In this work we investigate theoretically the surface energy, stress and work function of elastically deformed metal. An uni-axial strain applied to the surface introduces anisotropy to the metal by changing the density (or separation) of atomic planes, electron gas concentration, and contributes to an extra surface dipole barrier. A rigorous study of this problem from first principles is tedious and requires heavy numerical computations. On the other hand, the calculations based on the isotropic models of metal, i.e., on the jellium model [24], which ignores the discrete nature of ions, or the stabilized jellium model, in which interparticle interactions are averaged over volumes of the spherical Wigner-Seitz cells, do not allow to account properly for the effects of inhomogeneous strain. In this work we develop a modification of stabilized jellium model, in order to describe the metal deformed by the strain [25-27]. In this modification the metal energy is expressed as a function of the density parameter $r_s$ and of the given deformation. The following section provides a general discussion of the effect of deformation-induced anisotropy on work function which is one of the most important electronic surface characteristics. In Sec. III we present equations for stabilized jellium model accounting for elastic deformation. In Sec. IV the modified stabilized-jellium model is applied to calculate, by the Kohn-Sham method, the effect of uniaxial strain on electronic surface characteristics of single crystals of aluminum.

## II. THE DESCRIPTION OF DEFORMATION

It is important to note that in all experiments we have to do with the *finite* samples. A different reticular electron density at particular faces of single crystal (crystallite) of irregular shape leads to the difference of electrostatic potential for these faces. Analogous situation may be ascribed to take place for the deformed metal.

Let us consider a hypothetical crystal in the shape of rectangular parallelepiped (see Fig. 1). We assume the equivalence of all its faces in the undeformed state. This picture breaks down owing to the crystal deformation. The four side-faces remain equivalent one to the other, but not to the two base faces. The electroneutrality condition for the tensed or compressed metallic sample can be written in the form

$$\int dx \int dy \int dz [n(x,y,z) - \rho(x,y,z)] = 0 , \tag{1}$$

where the electron charge density distribution $n(\vec{r})$ attains the magnitude $n_0$ in the metal-bulk. The ionic charge distribution can be modeled by the step function,

$$\rho(\vec{r}) = \bar{\rho}\,\theta(\vec{r} - \vec{r}\,'),$$

where $\vec{r}\,'$ is the radius-vector of surface, $\bar{\rho} = \bar{n}_0 / Z$, and Z is the valence. We use atomic units ($e = m = \hbar = 1$) throughout.

By definition [14], putting the electrostatic potential in the vacuum equal zero, the electron work function for certain face of semi-infinite crystal is

$$W_{face} = -\bar{\phi}_0 - \frac{d}{dn}\left(\bar{n}_0 \varepsilon_J\right) - \left\langle \delta v \right\rangle_{face}, \qquad (2)$$

where $\bar{\phi}_0 < 0$ denotes the electrostatic potential in the metal bulk and $\varepsilon_J \equiv \varepsilon_J(\bar{n}_0)$ is the average energy per electron in the uniform electron gas. The last term, which represents the difference $\delta v(\vec{r})$ between the pseudopotential of the lattice of ions and the electrostatic potential of positive background averaged over the Wigner-Seitz cell, allows to distinguish between different faces of crystal (compare Sec. III).

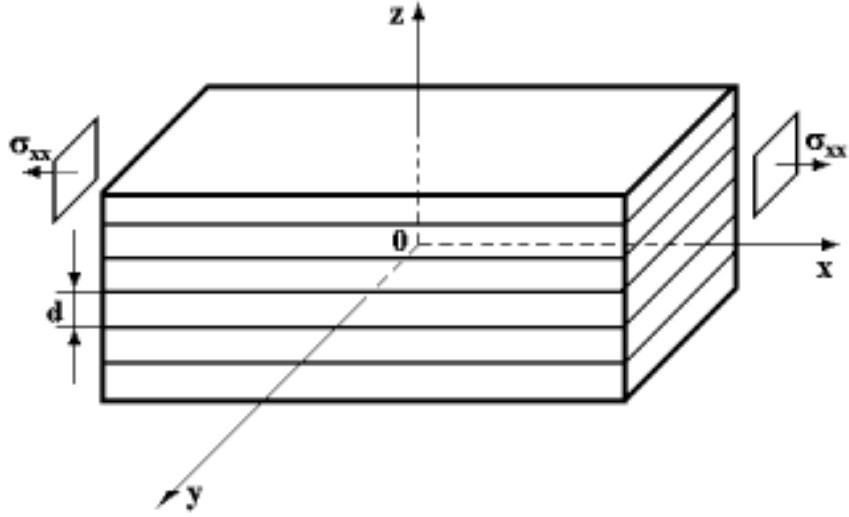

FIG.1. *The qualitative sketch of the sample deformation.*

For a deformed sample we assume *equivalence* of the *y*-and *z*-directions. The deformation along the *x*-axis induces an artificial homogeneous anisotropy. It seems, as if the work functions along the *x*- and *z*-direction were different for the finite sample. This conclusion is not correct. It is connected with a widely-distributed point of view (see [28] and references therein), that the work function "anisotropy" is determined by the reticular electron density of the given crystal face. However, the electron work function is defined as the difference between the electron energy level in the vacuum and at the Fermi surface. This difference is independent of space directions and coordinates and is constant for a metallic sample. The work function (or ionization potential) is a scalar quantity.

From the viewpoint of the finite sizes of a sample the considerations presented by Smoluchowski [28] and by Lang and Kohn [29] are correct in the case when all faces of a finite sample posses the same atomic packing density. For the cubic crystals it is a parallelepiped with all sides having equivalent Miller indices. In the case of a sample of arbitrary form, the work function depends, in general, on the orientation of all parts of the surface[1].

By the way, the ``imaginary'' difference of work functions along the $x$- and $z$-direction, $W_x - W_z$, defined by means of the habitual form (2), equals zero. It leads to a significant inequality

$$\overline{\phi}_x - \overline{\phi}_z = -\langle \delta v \rangle_x + \langle \delta v \rangle_z \neq 0 ,$$ (3)

which means that the values of electrostatic potential in the bulk of the metal, $\overline{\phi}_x$ and $\overline{\phi}_z$, could be treated as if they were corresponding to different semi-infinite crystals. This dramatic inequality does not allow to define unequivocally the work function of a finite macroscopic sample because the surface electrostatic barrier is different for different directions.

For simplicity of analysis let us express the electron profile of sample

$$n(\vec{r}) = n_0(\vec{r}) + \delta n(\vec{r}),$$ (4)

and

$$\overline{\phi} = \overline{\phi}_0 + \overline{\delta \phi} ,$$ (5)

where $n_0(\vec{r})$ and $\overline{\phi}_0$ are the values corresponding to semi-infinite metal. The ``surplus'' density $\delta n(\vec{r})$ originates from the electron transfer from one crystal side to another [31], and differs from zero only in the near-surface layer. Then, condition (1) along each direction has a trivial form

$$A_i \int_{-\infty}^{+\infty} dx [n_0(\vec{r}) - \rho(\vec{r})] = 0 ,$$ (6)

where $A_i \equiv A_x, A_y, A_z$ are the areas of faces of a macroscopic sample, and $A_y = A_z$. Taking into account (4) the Eq.(6) can be written in the ``cross-directional'' form

$$A_x \int_{-\infty}^{+\infty} dx \delta n(\vec{r}) + A_y \int_{-\infty}^{+\infty} dy \delta n(\vec{r}) + A_z \int_{-\infty}^{+\infty} dz \delta n(\vec{r}) = 0 ,$$ (7)

where the surplus charge at each side is proportional to its area. Here, for simplicity of illustration, we assume a constancy of $\delta n(\vec{r})$ at every side. From Eq.(7) follows the relation

$$\int_{-\infty}^{+\infty} dz \delta n(\vec{r}) / \int_{-\infty}^{+\infty} dx \delta n(\vec{r}) = -A_x / 2A_z , \qquad (8)$$

which means that the charges on these sides have the opposite sign. As a whole the sample must be neutral[2]. Corresponding changes of the electrostatic potential are defined by the Poisson equation, and yield the relations for the $x$ and $z$ components, of the identical form

$$\delta \overline{\phi}_x = -4\pi \int_{-\infty}^{+\infty} dx x \delta n(\vec{r}) = -C_x x_0 , \qquad (9)$$

where $x_0$ and accordingly $z_0$ are the positions of *self-induced* charge density at the lateral and base sides, and $C_x, C_z$ are constants. It enables us to speak about the appearance of additional, three-dimensional surface dipole barrier. By inequality

$$A_x << A_z , A_y \qquad (10)$$

(see Eq.(8)), we have for the weight coefficients $\left| C_x / C_z \right| \propto A_z / A_x$ and for the additional potentials $\left| \delta \overline{\phi}_y \right| = \left| \delta \overline{\phi}_z \right| << \left| \delta \overline{\phi}_x \right|$. Using (5), equation (3) can be rewritten as

$$\delta \overline{\phi}_x \approx \langle \delta v \rangle_z - \langle \delta v \rangle_x , \qquad \text{and} \qquad \delta \overline{\phi}_z = \delta \overline{\phi}_y \approx 0 . \qquad (11)$$

The condition (10) means that the work function is weakly dependent on the electron transfer between the faces perpendicular to $y$- and $z$-direction and thus, the measurement of work function at these faces may be replaced by the measurement for a semi-infinite metal. The *true* work function may be measured by the Kelvin method in the areas that are near the edges. These areas correspond to the change of sign of the density, $\delta n(\vec{r}) \approx 0$. For the photoemission method of measurements of work function, the conditions (10) and (11) mean that the registration of measured electrons should take place at the distances much greater than the linear dimensions of a sample. Otherwise, if the photon energy is not high enough, the electron escaping from a metal is not going to ``infinity'' but it may transit from one face onto the other.

The amount of surplus charge $Q_x$ transferred from one face to the other (see Eq.(9)) can roughly estimated with the help of ordinary electrostatic relation, $\delta \overline{\phi}_x \approx Q_x / \sqrt{A_x}$. Representing $A_x \approx N_x 2\pi r_0^2$, where $N_x$ is the number of the surface Wigner-Seitz cells of the radius $r_0$, we get $Q_x \approx 3r_0 \sqrt{N_x} \delta \overline{\phi}_x$. The condition $Q_x > 0$ means that $Q_x$ electrons are transfered from the base faces to the lateral ones. Thus, the surface energy per unit area changed by $-W_x Q_x / A_x$ and $+W_z Q_x / 2A_z$ at the base and lateral sides, respectively. A ratio of these values corresponds to (7). Here, $W_x Q_x$ equals to the work needed to remove $Q_x$ electrons from the base side of metallic sample to infinity and $W_i$ is the work function of given side $i$. Thus, self-charging of the surface may have an influence on anisotropy of surface energy of single crystal. For example, for an aluminum sample with $\delta \bar{\phi}_z \approx 0.5$ eV, and $N_x = 10^2, 10^4$, the electronic charge $Q_x \approx 1,10$, respectively. It is worth noting that for a small crystal (cluster) this charge may be very significant [34]. Therefore the elasticity and self-charging effects may play an important role in explanation of recently observed force and conductance fluctuations in the tensed metallic nanowires [35,36].

On the ground of the above discussion, and owing to Eq.(11), the properties of a large surface plane of deformed metallic crystal can be calculated in a standard manner.

### III. MODEL OF UNIFORMLY DEFORMED METAL

The dependence of the CPD on the uniaxial deformation $u_{xx}$ was measured for polycrystalline tensed samples [4,5]. We assume that deformation is a measured quantity and a polycrystal is to be considered as assembled from a number of simple crystallites. Thus, qualitatively, the problem can be reduced to the consideration of tension or compression applied to a single crystal.

Let us first express the average electron density in a metal as a function of deformation. In this purpose, consider an undeformed cubic cell of side length $a_0$ and volume

$$\Omega_0 = a_0^3 = \frac{4}{3}\pi r_0^3, \tag{12}$$

where $r_0 = Z^{1/3} r_s$ is the radius of spherical Wigner-Seitz cell. For uni-axially deformed cell, elongated or compressed along the $x$-axis, one can write

$$\Omega = a_x a_y^2 = \frac{4}{3}\pi a b^2, \tag{13}$$

where $a_x$ and $a_y = a_z$ are the sides of elementary parallelogram, $a$ and $b$ the half-axes of the equivalent prolate or oblate spheroid of revolution relative the $x$-axis. We also have

$$a_x = a_0(1 + u_{xx}), \quad \text{and} \quad a_z = a_0(1 + u_{zz}) = a_0(1 - \nu u_{xx}), \tag{14}$$

where $\nu$ is the Poisson coefficient for polycrystal, and $\Omega / \Omega_0 - 1 = u_{xx} + u_{yy} + u_{zz}$. From (12)-(14) follows

$$a = r_0(1 + u_{xx}), \quad \text{and} \quad b = r_0(1 - \nu u_{xx}), \tag{15}$$

Similarly, the spacing between the lattice planes perpendicular to the $y$- or $z$-directions is

$$d_u = d_0 (1 - \nu u_{xx}),$$
(16)

where $d_0$ is the interplanar spacing in undeformed crystal. Consequently, following (12)-(15), the average electron density in deformed metal is given by

$$\bar{n} = \bar{n}_0 \Omega_0 / \Omega = \bar{n}_0 [1 - (1 - 2\nu) u_{xx}] + O(u_{xx}^2),$$
(17)

and the corresponding density parameter is

$$r_{su} = r_s [1 + (1 - 2\nu) u_{xx}]^{1/3}.$$
(18)

Proceeding similarly as by the derivation of equations for the original stabilized jellium model [25], we consider a metal assembled from Wigner-Seitz cells. The average energy per valence electron in the bulk is

$$\varepsilon = \varepsilon_J (\bar{n}) + \varepsilon_M + \overline{w}_R,$$
(19)

where the first term gives the jellium energy

$$\varepsilon_J (\bar{n}) = \frac{3 k_F^2 (\bar{n})}{10} - \frac{3}{4\pi} k_F (\bar{n}) + \varepsilon_{cor} (\bar{n}),$$
(20)

consisting of average kinetic and exchange-correlation energy per electron, $k_F = (3\pi^2 \bar{n})^{1/3}$. The remaining two terms in (19) represent the average of the repulsive part of the Ashcroft model potential, the Madelung energy. A small, band structure energy term [25,37] is neglected in (19).

By transformation of ordinary jellium into stabilized one, the Coulomb interactions were averaged, as usually for isotropic medium, over the Wigner-Seitz cells. The uni-axial strain applied to the crystal deforms the spherical Wigner-Seitz cells into ellipsoidal ones. This has an influence on the Madelung energy, $\varepsilon_M$, which now should be averaged over the volume of the deformed cell. It can be expressed in the form similar to the gravitational energy of the uniform spheroid [38] to give

$$\varepsilon_M (\bar{n}) = \frac{1}{Z} \int\limits_{spheroid} d\Omega \bar{n} \left[ -\frac{Z}{r} \right] + \frac{1}{2Z} \int\limits_{spheroid} d\Omega \bar{n} V(r)$$

$$= \begin{cases} -\dfrac{9Z}{10a} \dfrac{1}{2p} \log \dfrac{1+p}{1-p}, a > b, \\[4mm] -\dfrac{9Z}{10a} \dfrac{1}{p} \arctan p, b > a, \end{cases}$$
(21)

where $V(r)$ is the electrostatic potential inside an uniformly charged spheroid, $p = \sqrt{\left| 1 - b^2 / a^2 \right|}$ defines spheroid's eccentricity, and the upper/lower form corresponds to prolate/oblate spheroid, respectively. This expression has a correct limit: $\varepsilon_M(\overline{n}) \rightarrow -0.9 Z / r_0$ for $u_{xx} \rightarrow 0$.

We assume that the shape of ionic cores is not influenced by the deformation and remain spherical, thus $\overline{w}_R = 2\pi \overline{n} r_c^2$. For the potential difference $\delta v(\vec{r})$, averaged over the Wigner – Seitz cell [25], holds the same relation as for an undisturbed crystal

$$\langle \delta v \rangle_{WS} = \widetilde{\varepsilon} + \varepsilon_M + \overline{w}_R, \tag{22}$$

where electrostatic self- energy of the uniform negative background inside a spheroid is

$$\widetilde{\varepsilon} = -\frac{2}{3} \varepsilon_M. \tag{23}$$

The pseudopotential core radius can be found from the condition of mechanical equilibrium depending on the mechanical stress induced in the volume of cell. In order to determine the core radius $r_c$ let us note that for the strained metal the intrinsic pressure in the bulk of metallic sample, $P = -dE / d\Omega = \overline{n}^2 d\varepsilon / d\overline{n}$ is compensated dy the pressure exerted by external forces,

$$P = -\left( \sigma_{xx} + \sigma_{yy} + \sigma_{zz} \right) = -Y u_{xx} (1 - 2\nu), \tag{24}$$

where $\sigma_{ii}$ are the components of the tensor of mechanical stress and $Y$ is the Young's modulus.

Thus, for a strained metal, the averaged energy per electron in the bulk is

$$\varepsilon = \varepsilon_J(\overline{n}) + \varepsilon_M + \overline{w}_R + P / \overline{n}. \tag{25}$$

For ideal metal $\nu = 1/2$ and $P = 0$. It means that external force changes not volume but a shape of cell or sample. In the linear approximation Madelung energy (21) is good approximated by $\varepsilon_M(\overline{n}) \rightarrow -0.9 Z / r_{0u}$. Inserting the explicit expressions for (20), (21) and (24) into (25), from the minimum condition we have

$$r_c = \left\{ -\frac{2}{15} \left( \frac{9\pi}{4} \right)^{2/3} r_S + \frac{1}{6\pi} \left( \frac{9\pi}{4} \right)^{1/3} r_S^2 + \frac{1}{5} Z^{2/3} r_S^2 + \frac{2}{9} r_S^4 \frac{d\varepsilon_{cor}}{dr_S} + \frac{8}{9} \pi r_S^6 P \right\}_{r_S = r_{su}}^{1/2}, \tag{26}$$

where $r_{su}$ is the equilibrium density parameter of the strained metal. Here, we assume that the volume of spheroid is equal to the volume of equivalent sphere of radius $r_{0u} = Z^{1/3} r_{su}$. Since

$$\langle \vec{\vec{a}} v \rangle_{WS} = \overline{n} \frac{d}{d\overline{n}} \left( \varepsilon_M + \overline{w}_R \right), \tag{27}$$

then at the equilibrium density for the strained metal we get

$$\langle \delta v \rangle_{WS} = -\overline{n}\frac{d}{d\overline{n}}\left[\varepsilon_J(\overline{n}) + \frac{P}{\overline{n}}\right]. \tag{28}$$

Subsequently, similarly as Perdew et al. [25] we can introduce the face-dependence of the stabilization potential

$$\langle \delta v \rangle_{face} = \langle \delta v \rangle_{WS} - \left(\frac{\varepsilon_M}{3} + \frac{\pi\overline{n}}{6}d_u^2\right). \tag{29}$$

The total energy of finite crystal may be written as the sum of the bulk $E^b$ and surface energy $E^s$, where

$$E^S = \gamma_y 4A_y + \gamma_x 2A_x. \tag{30}$$

Here, $\gamma_y$ and $\gamma_x$ are surface energies, per unit area, of the lateral and base sides, respectively. In the undeformed state $\gamma_x = \gamma_y = \gamma_z \equiv \gamma$ and surface energy (30) changes by

$$dE^S = 4A_y\left(\gamma\delta_{\alpha\beta} + \frac{d\gamma}{du_{\alpha\beta}}\right)du_{\alpha\beta} + 2A_x\left(\gamma\delta_{\alpha\beta} + \frac{d\gamma}{du_{\alpha\beta}}\right)du_{\alpha\beta}, \tag{31}$$

where $\alpha$ and $\beta$ denote directions in the plane of lateral and base sides and $\delta_{\alpha\beta}$ is the Kronecker symbol. Following our model we calculate only

$$\tau_{xx} = \gamma + \frac{d\gamma}{du_{xx}}. \tag{32}$$

The work function is calculated from the displaced – profile change-in-self-consistent field (DP$\Delta$SCF) expression instead of Eq. (2).

For a discussion of our results it is useful to rewrite Eq.(2) in the following form

$$W_{face} = -\overline{v}_{eff} - \varepsilon_F, \tag{33}$$

where the effective potential in the bulk, $\overline{v}_{eff} = \overline{\phi} + \overline{v}_{xc} + \langle \delta v \rangle_{face}$, gives the total barrier height at the metal-vacuum interface, and $\overline{v}_{xc}$ is the exchange-correlation potential in the bulk ( $\overline{v}_{xc} = v_{xc}(-\infty)$ ),

$$v_{eff}(z) = \begin{cases} \phi(z) + v_{xc}(z) + \langle \delta v \rangle_{face}, & z < 0, \\ \\ \phi(z) + v_{xc}(z), & z > 0, \end{cases}$$

## IV. RESULTS AND DISCUSSION

To verify the theory presented in Sec. III, the Kohn-Sham equations were solved for two most densely packed surfaces of Al represented by the stabilized jellium model. In the language of our model, we consider two regular single crystals of Al which have all their sides eguivalent in undeformed state. Owing to the crystal deformation the four side-faces remain equivalent one to the other, but not to the two base faces (Fig.1). The $\left\langle \delta v \right\rangle_{face}$ term included into the effective potential allows to generate the face-dependent density profiles which were used to calculate the surface characteristics: work function, surface energy and surface stress. All calculations were carried out for the upper side of the sample (see Fig.1) assuming the polycrystalline value of the Poisson coefficient $v = 0.36$ for elastic properties of Al [39].

Within the applied range of deformation, $-0.03 \leq u_{xx} \leq +0.03$, the changes in surface quantities remain linear. The positive/negative deformation $u_{xx}$ means tension/compression of the side of a sample, i.e., the decrease/increase of the atomic packing density at this side, and the decrease/increase of the mean electron concentration $\bar{n}$ and interplanar spacing in the direction perpendicular to the concidered crystal side. For better understanding of the crystal effects we have also performed calculations for the special case of "ideal" metal for which $v = 1/2$. In this case the deformation does not change $\bar{n}$, however, the second term (corrugation dipole barrier) in the face-dependent potential (29) will be changed.

The results of calculations are summarized in Table 1. As is seen the surface energy increases linearly with the applied positive deformation $u_{xx}$ and decreases with the negative one. It means that $d\gamma / du_{xx}$ is positive both for $u_{xx} > 0$ and for $u_{xx} < 0$. Accordingly, Eq. (32) gives the values of the component of surface stress, $\tau_{xx}$, lager than surface energy. For $u_{xx} > 0$ surface stress is somewhat lager than for $u_{xx} < 0$. Let us consider the case of "ideal" metal, $v = 1/2$. It seems that ideal metal fits better to the classical definition of surfase stress [19,20]. This is connected with the fact that subjected to deformation ideal metal changes only its surface area – the electron concentration in its bulk remains unchanged. The calculations performed for Al (111) surface yield the values of the strain derivative $d\gamma / du_{xx} = 247$ and 213 erg/cm$^2$ for $u_{xx} > 0$ and $u_{xx} < 0$, respectively. These values are much smaller then the ones reported in Table 1. In this case $(v = 1/2)$ we can also evaluate the other components of surface stress $\tau_{zz} = \tau_{yy} = \gamma + d\gamma / du_{yy}$. Substituting $du_{zz} = du_{yy} = -v du_{xx}$ we get $\tau_{zz} = \tau_{yy} = \gamma - 2d\gamma / du_{xx} < \gamma$. Let us make two observations at this point. First, the latter result agrees with our estimation $(\tau < \gamma)$ that we have presented in Sec.I, and the results derived on the basis of the elasticity theory [40] where the $\tau / \gamma$ ratio was expressed in terms of the Poisson coefficient, $\nu$, to give $(3\nu - 1)/(1 - \nu)$. For $\nu = 1/2$, this formula gives $\tau/\gamma = 1$ and $\tau < \gamma$ for $\nu < 1/2$. Second, in order to calculate $\tau_{zz}$ and $\tau_{yy}$ for a sample tensed (compressed) along the $x$- axis we should exploit $d\gamma/du_{xx}$ for $u_{xx} < 0$, whereas for a compressed sample, the corresponding value for $u_{xx} > 0$ ($u_{xx} < 0$). This is because the tension applied along $x$-direction causes that a sample is compressed along the orthogonal (y and z) axes. Calculated surface stress for Al(111) is in a very good agreement with the values resulting from the available *ab initio* calculations: 1441 erg/cm$^2$, Ref.[15], and 1249 erg/cm$^2$, Ref.[41]. It gives also improvement over the results obtained for ordinary jellium [24,41] and previous direct application of the stabilized-jellium model [13].

TABLE I. *Calculated surface energies,* $\gamma$, *work function W, strain derivative* $d\gamma/du_{xx}$ *and surface stress,* $\tau_{xx}$, *for elastically deformed Al ($r_s = 2.06$) samples ($u_{xx} = \pm 0.03$, positive and negative deformation are labeled with (+) or (-)).* $\Delta W$ *is the work function difference. The values of Young's modulus are: 70 GPa (Al).* $\Delta \overline{v}_{eff}(z_0)$ *is the shift of effective potential beyond.*

| Metal | Face | $\gamma$ (erg/cm$^2$) | W (eV) | $u_{xx}$ | $d\gamma/du_{xx}$ (erg/cm$^2$) | $\tau$ (erg/cm$^2$) | $\Delta W$ (eV) | $\Delta \overline{v}_{eff}(z_0)$ (eV) |
|-------|------|------|------|------|------|------|------|------|
| Al | (111) | 946 | 4.096 | (+) (-) | 460 400 | 1406 1346 | -0.032 +0.033 | -0.103 +0.106 |
| Al | (100) | 1097 | 3.780 | (+) (-) | 833 810 | 1930 1907 | -0.025 +0.016 | -0.064 +0.069 |

The work function decreases linearly with $u_{xx}$ but the relative change is less than 1% (see Table I) for the considered strains. The similar behaviour is observed for $\nu = 1/2$. It is seen that the dominating component, which leads to a decrease of W with $u_{xx}$, is a change in the $\langle \delta v \rangle$ term. Thus, the change of work function under the influence of deformation is determined by the competition of negative change both in the exchange-correlation, $v_{xc}$, and electrostatic, $\phi$, components of the effective potential $v_{eff}$ and the positive change in the face-dependent component $\langle \delta v \rangle_{face}$. A dominant role is played by the change of $\langle \delta v \rangle_{face}$ term while the change in the Fermi energy is quite unnoticeable. An overall decrease/increase in the work function W is determined by a positive/negative shift of the electrostatic potential in the metal interior.

The calculated change of the work function with strain seems to contradict the experimental results [3-6] where it was found that work function increases/decreases with elongation/compression of the sample. This conclusion was based on the analysis of the measured CPD [3-7,9,27]. In the following we demonstrate that this contradiction is spurious. The point is that the measurement by Kelvin method fixes the change of surface potential. So, the explanation of experimental observations can be given basing not on the change in the work function but analyzing the change in the effective potential, $v_{eff}$, upon deformation. The Kelvin method gives the value of the potential difference at the surface of a sample which one can define as the position of the image plane $z = z_0$ [26]. In distinction to the work function, to which $\langle \delta v \rangle_{face}$ contributes directly (Eq.(2)), at the image-plane position which is located outside the geometric surface, the effective potential feels the change in $\langle \delta v \rangle_{face}$ by means of the self-consistent procedure in solving Kohn-Sham equations (notwithstanding $\langle \delta v \rangle_{face}$ is nonzero inside a sample only). The calculations performed for Al(111) demonstrate that, the ratio of the effective potential difference, $\Delta v_{eff}$, between strained ($u_{xx} = \pm 0.03$) and strain-free samples, at the surface and in the bulk, is $\Delta v_{eff}(z = z_0)/\Delta \bar{v}_{eff} \approx -3$ (see Table 1). Here, $\Delta \bar{v}_{eff}$ denotes the respective difference in the metal bulk. This ratio dmonstrates our main conclusion on the difference between CPD and work function.

The results for $\Delta v_{eff}(z_0; u_{xx})$ are shown also in Table I. The potential difference outside the sample is more negative as deformation increases. The calculated changes in the effective potential have the same sign as the measured CPD for Al. For a polycrystalline Al sample subject to deformation $u_{xx} = 0.03$, the CPD amounts $-0.025 \pm 0.002$ Volts [5]. Since a polycrystalline sample can be considered as assembled from arbitrarily oriented single crystals, the values obtained by us should be averaged in order to compare them with experiment. So, both experiment and calculations give a negative change of surface potential,

$$CPD = \Delta v_{eff}(z = z_0) < 0.$$

For the conventional method of measurement of the work function changes upon strain [4,5,9] this means:

$$W(u_{xx}) = W(0) - CPD(u_{xx}) > W(0),$$

i.e., the work function increases for a tensed sample. Thus, on the whole our results agree with the independent experiments both for tensed [4-6] and compressed [1,3] metallic samples. The results for $\Delta v_{eff}(z_0, u_{xx})$ correspond to direct observation of stress-induced shift in the measured contact potential: the effective potential outside the open faces of a sample is more negative/positive when tensile/compressive force is applied. However, unlike the effective potential at the surface, due to the different effect of the $\langle\delta v\rangle_{face}$ term the value of the potential in the metal bulk is more positive/negative for an expanded/compressed sample. So, for Al sample the work function change vs. strain shows an opposite trend compared to that of contact potential which differs also from that predicted by non self-consistent calculations [27]. Accordingly, the results of Table I demonstrate that work function decreases with $u_{xx}$. In other words, our results show that the measurements by Kelvin method give not a change in the work function upon strain but a change in the surface potential.

In summary, the stabilized-jellium model has been extended to encompass the effects of elastic strain on surface properties of simple metals. By imposing uniaxial strain to metal surface and limiting ourselves to linear terms in deformation, we have obtained a realistic description of strain dependence of surface quantities: surface energy, surface stress and work function. We have presented a consistent explanation of experiments on stress-induced contact potential difference at metal surfaces.

This work has been supported by the grant of Institute of Experimental Physics Wroclaw University and NATO "Science for Peace" Programme (project SfP-974109).

---

[1] In special case of nonzero quadrupole moment of charge distribution in elementary cell the effective potential in the bulk depends on the shape of a sample [30].

[2] It should be noted that the phase shift $\eta_k$ of the single-particle wave function along each direction depends on the potential shape in the vicinity of the surface and the Sugiyama-Langreth neutrality sum-rule [32] must be rewritten with taking into account the anisotropy (i.e., the self-charging) [33].